\documentclass[twoside]{ilcws07}
\usepackage[latin1]{inputenc}
\usepackage[dvips]{graphicx,epsfig,color}
\usepackage{wrapfig,rotating}
\usepackage{amssymb,amsmath,array}

\begin{document}
\title{
%%%%   Paper title goes here  %%%%%%%%%%%%%%
ILC Positron Production Target Simulation } %% 
%***********************************************************************
% AUTHORS INFORMATION AREA
%***********************************************************************
\author{V.Gharibyan
\vspace{.3cm}\\
% Addresses and institutions (remove "1- " in case of a single institution)
 DESY - MDI \\
Hamburg - Germany
%% Remove the next three lines in case of a single institution
}
%%***********************************************************************
% END OF AUTHORS INFORMATION AREA
%***********************************************************************

\maketitle

\begin{abstract}
 A photon-positron conversion target 
of the undulator or laser based polarized positron source
is optimized using a modified GEANT-3 program adapted to count 
the spin transfer.  
High intensity positron beam with around 0.75 polarisation could be achieved choosing 
tungsten conversion target of 0.3 and 0.7 radiation lengths for the undulator and laser 
case respectively. 
\end{abstract}

\section{Positron Sources}
Currently two scenarios are considered to generate polarized positrons for the ILC.
Both are utilizing low energy circularly polarized photons and high 
energy electrons to boost these photons to MeV energies and then convert them 
into electron-positron pairs. Each method named after the photon source as 
undulator~\cite{Alexander:2003fh} 
or laser~\cite{Sakaue:2007zr} based positron production.
%\begin{wrapfigure}{r}{0.5\columnwidth}
\begin{figure}[htp]
\includegraphics[scale=0.39]{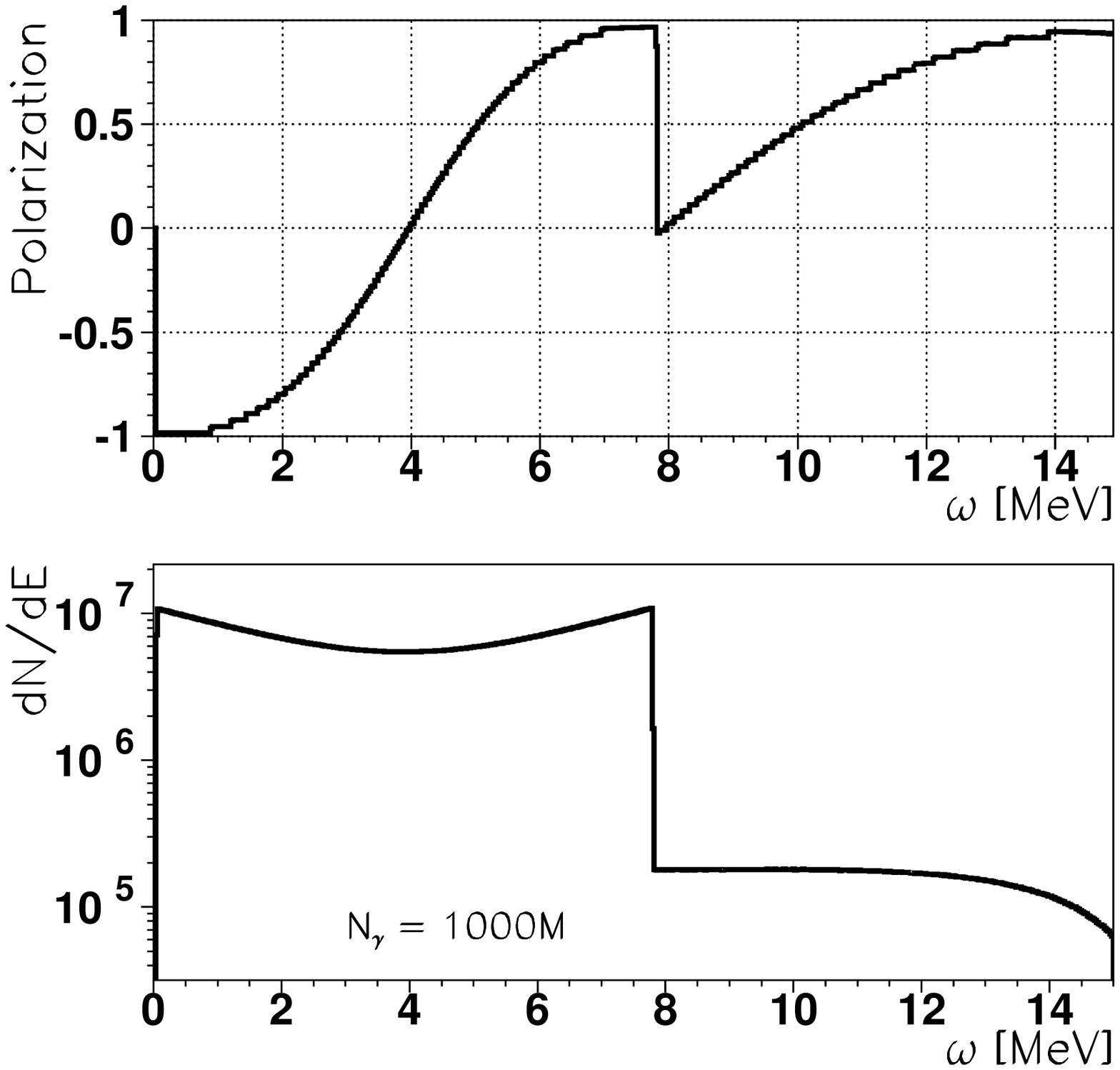}
\includegraphics[scale=0.39]{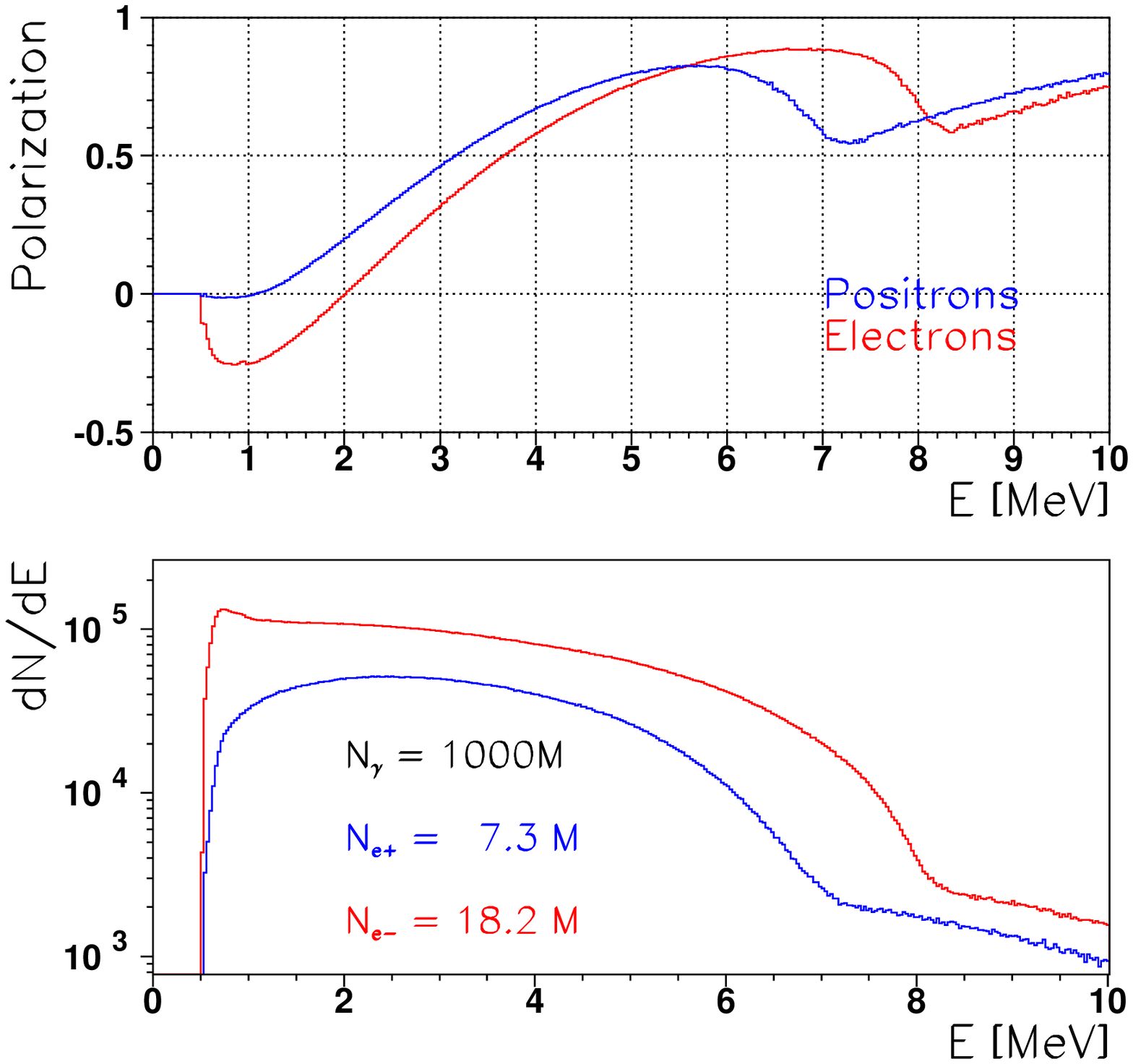}
\caption{Polarisation distribution and spectrum of undulator photons (left) and
positrons/electrons behind $0.2X_0$ tungsten (right). }
\label{Fig:1}
\end{figure}
%\end{wrapfigure}
 In this study~\cite{url} we will vary thickness and material 
of the production target to optimize the positron yield and polarisation. 

\section{Simulation Tools and Considered Polarized Processes}
\begin{figure}[htp]
\includegraphics[scale=0.39]{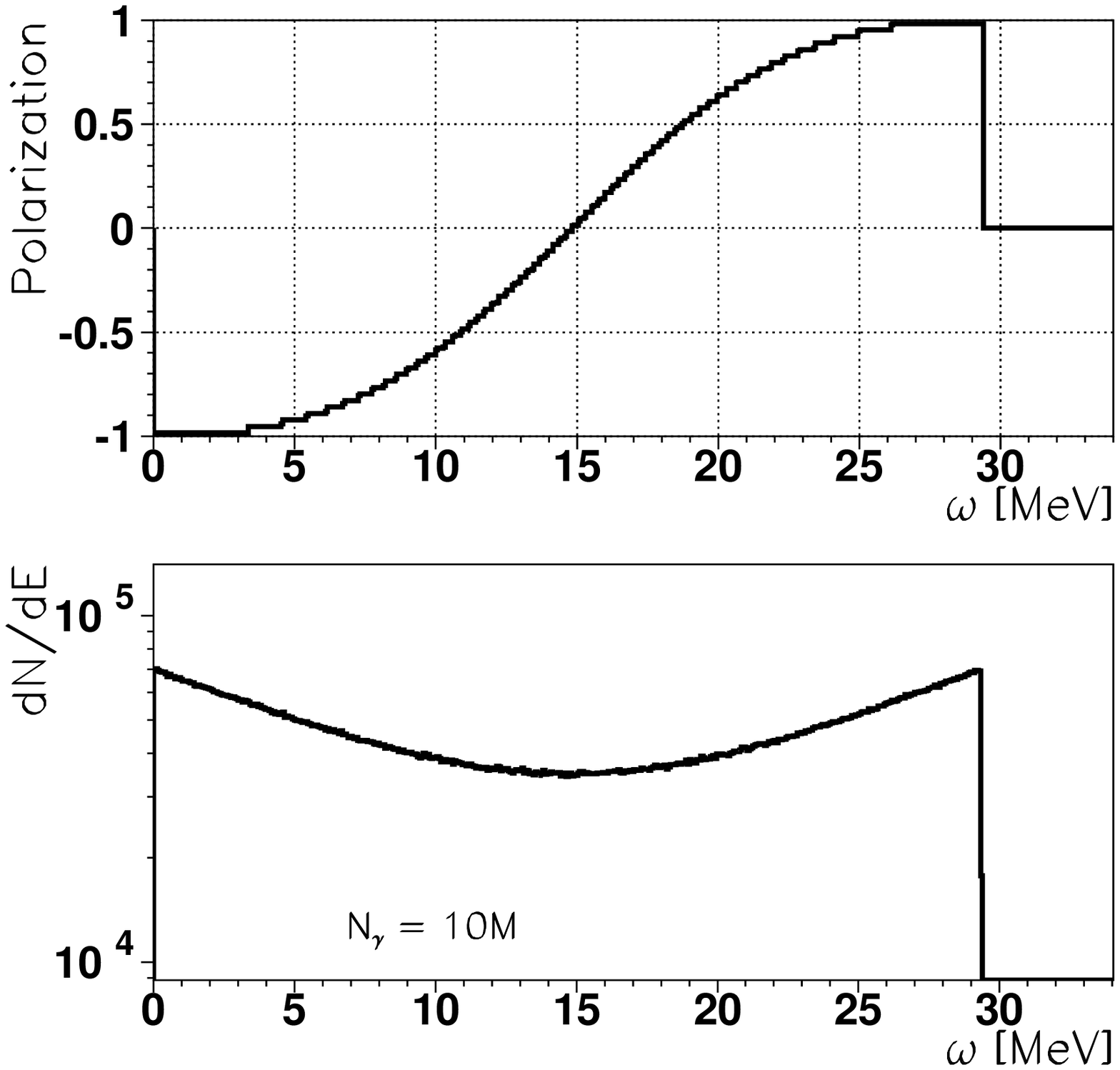}
\includegraphics[scale=0.39]{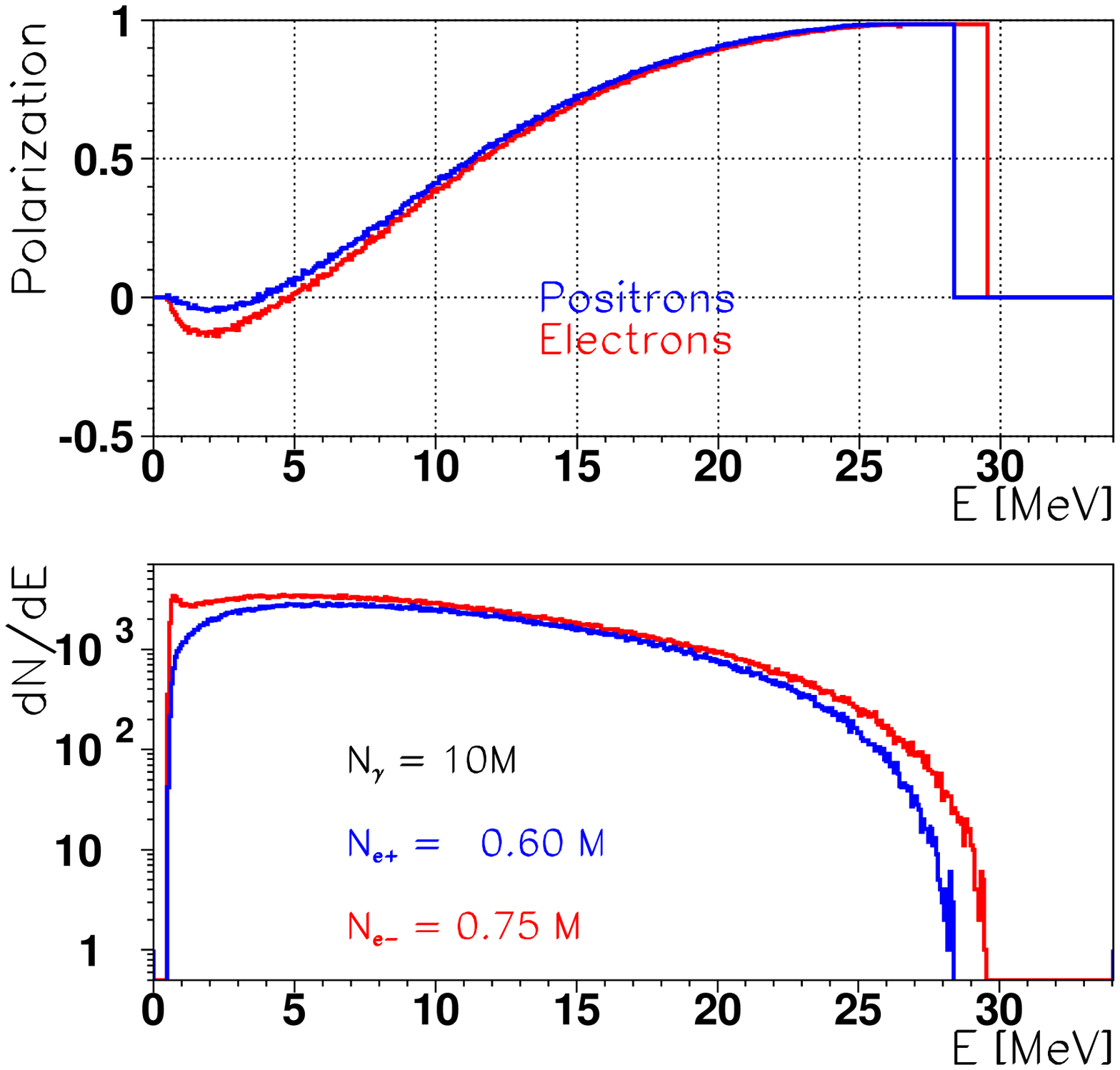}
\caption{The same as in Figure~\ref{Fig:1} for a laser scattering Compton photons and
$0.6X_0$ tungsten target }
\label{Fig:2}
\end{figure}
To achieve high number of positrons the target should be thick, of the order 
of one radiation length, $1X_0$ hence, the MeV photons may initiate showers, or at 
least 2-3 generations of particles and  
we would need a proper tracking tool like EGS or GEANT.   
\begin{wraptable}{r}{0.5\columnwidth}
%\begin{table}
%\centerline{\begin{tabular}{|l|l|}
\begin{tabular}{|l|l|}
\hline
Process  & Particles    \\
\hline\hline
&\\
Pair creation & $\vec{\gamma}\rightarrow\vec{e^+}\vec{e^-}$ \\
Compton scattering & $\vec{\gamma}+{e^-}\rightarrow\vec{\gamma}\vec{e^-}$  \\
Photoeffect &$\vec{\gamma}+{e^-}\rightarrow\vec{e^-}$ \\
Annihilation & $\vec{e}+{e^-}\rightarrow\vec{\gamma}$  \\
Bremsstrahlung & $\vec{e}+N\rightarrow\vec{e}+N+\vec{\gamma}$ \\
Multiple Scattering & $\vec{e}+N\rightarrow\vec{e}+N$   \\
Energy loss dE/dX & $\vec{e}+Ne\rightarrow\vec{e}+Ne^*$  \\
\hline
\end{tabular}
\caption{Modified GEANT-3 processes with polarisation transfer. Vector sign 
indicates polarized particle}
\label{tab1}
\end{wraptable}
To address the polarisation issue one have to take care also about the
polarisation tracking. This has been incorporated into the EGS by 
K.~Fl{\"o}ttmann~\cite{floettmann} and into the GEANT-4 by DESY-Zeuthen
E166 group~\cite{Geant4:PhysRefMan}. Here we will use GEANT-3~\cite{g3manual}
modified in a way to account polarisation transfer in the 
processes summarized in table~\ref{tab1}.
 For the multiple scattering and $dE/dX$ energy loss continuous approximations
there are certain  difficulties to treat the depolarisation especially for energies below
the critical $E_C$ and this is mostly because lack of the theoretical and 
experimental input. Anyhow, for our calculations we use straight trajectory/no depolarisation
for the $dE/dX$ loss and $(q+\cos{\theta})/(1+q\cos{\theta})$ approximation as depolarisation
factor 
for a $\theta$ multiple scattering angle with $q=(\gamma^2 -1)/(\gamma^2 +1)$ where $\gamma$
is the Lorentz factor.  

%While the first five are discrete and relatively well defined processes,
%the last two are continuous approximations and deserve closer
%attention especially for polarisation transfer or depolarisation.  
\section{Results}
One example of the simulation outcome for a $0.2 X_0$ tungsten is shown on 
Figure~\ref{Fig:1}
where initial photons originate from a helical 1m long undulator with 
k=0.19 on a 46.6~GeV energy electron beam (E166 experiment configuration). 
On the lower right figure one can find also total number of the 
positrons/electrons for the $10^9$ simulated initial photons. 
Figure~\ref{Fig:2} displays results for the initial Compton photons produced 
by a $532~nm$ laser, scattered on 1.3~GeV electrons with a crossing angle 
of 8~deg. Number of simulated Compton events is $10^7$.   
Using positrons intensity $dN/dE$ and polarisation $P$ distributions we can 
\begin{wrapfigure}{r}{0.5\columnwidth}
\includegraphics[width=0.5\columnwidth]{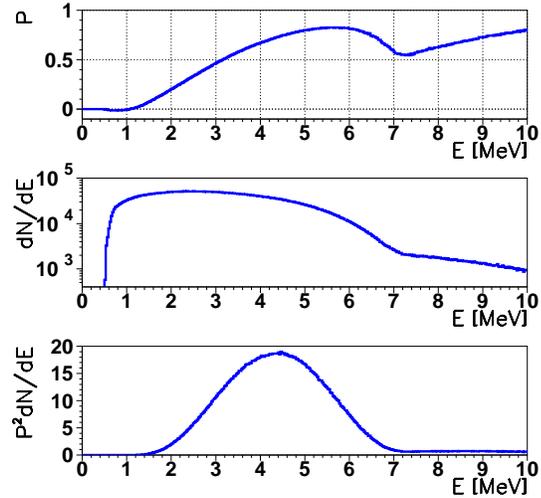}
\caption{Figure of merit derived for the $0.2X_0$ W case. (Figure~\ref{Fig:1}).}
\label{Fig:3}
\end{wrapfigure}
form a product $P^2 dN/dE$ (Figure~\ref{Fig:3}) to serve as a figure of merit 
for the target material and thickness optimization.

\section{Optimal Target}
\begin{figure}[htp]
\includegraphics[scale=0.39]{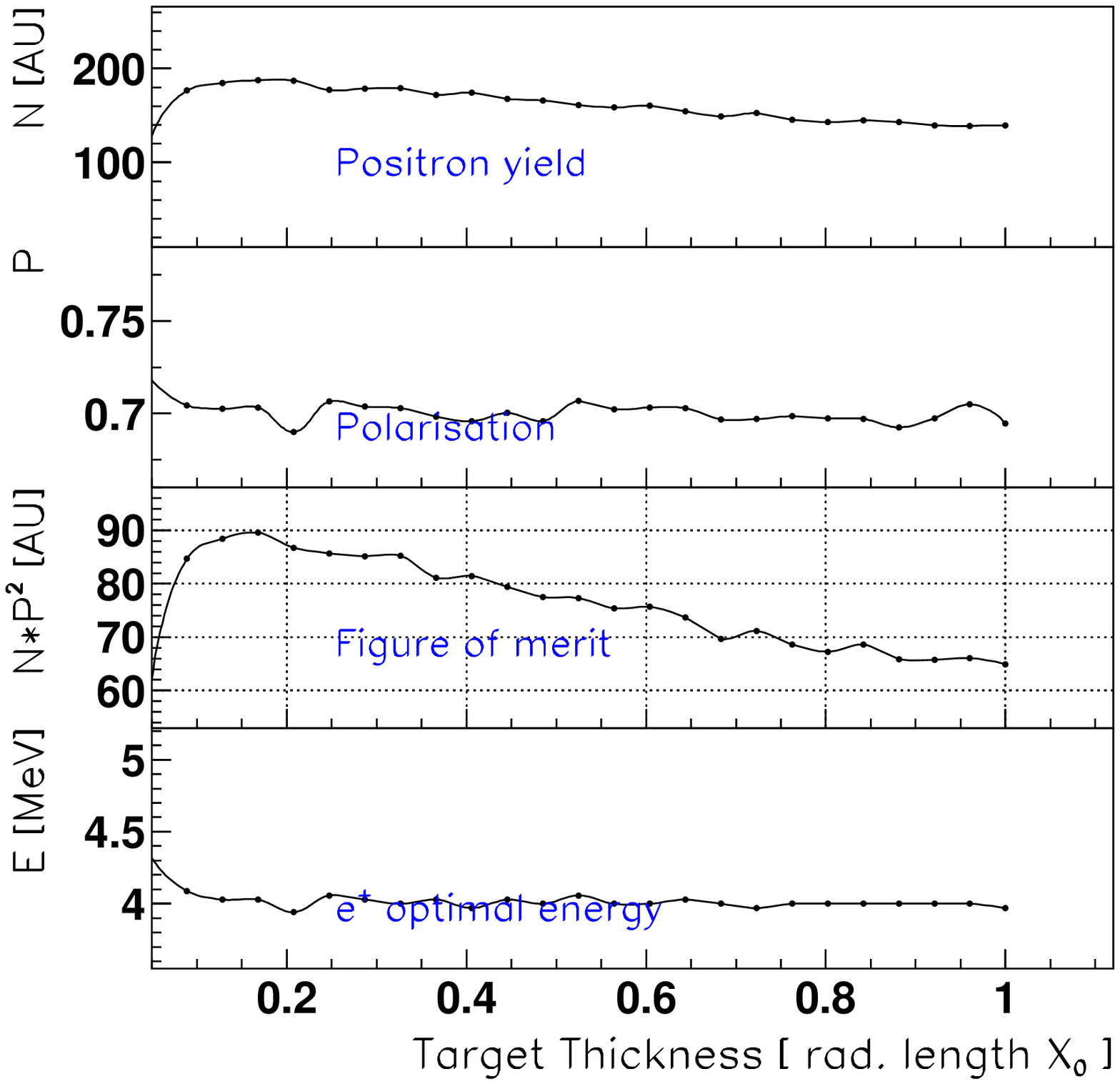}
\includegraphics[scale=0.39]{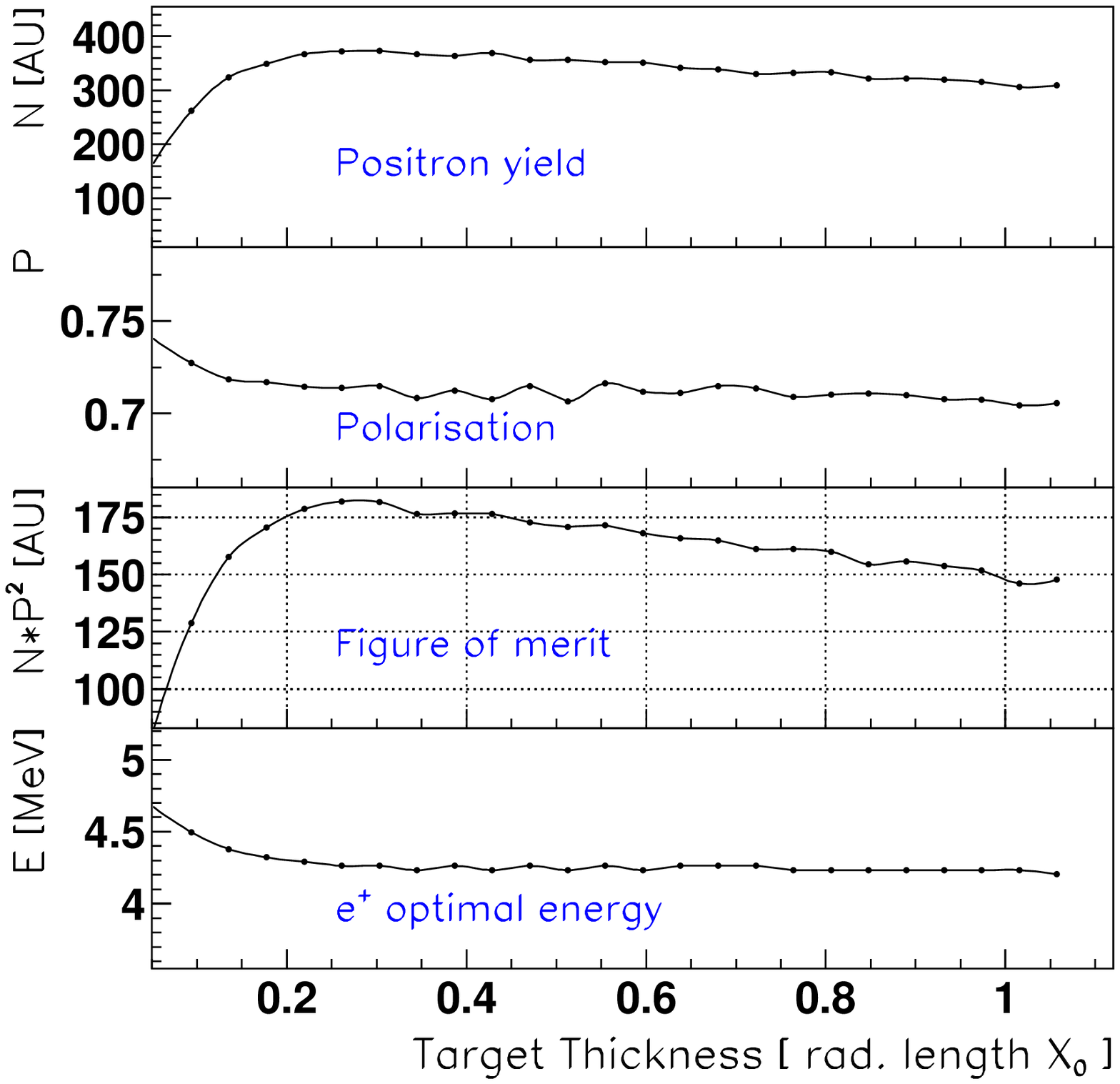}
\caption{Undulator produced positrons intensity, polarisation, figure of merit
and energy versus titanium (left) and tungsten (right) target thickness.}
\label{Fig:4}
\end{figure}

\begin{figure}[htp]
\includegraphics[scale=0.39]{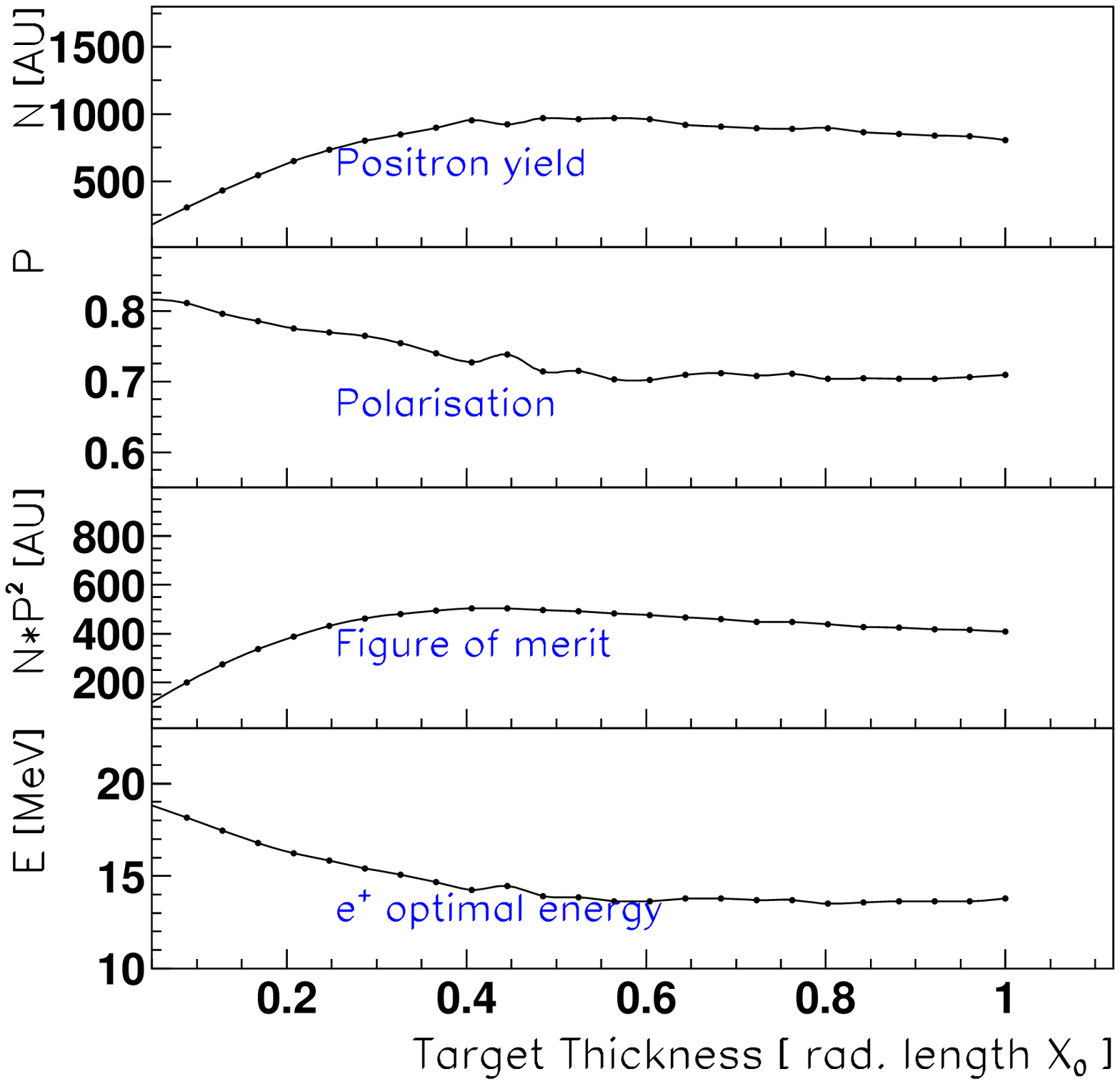}
\includegraphics[scale=0.39]{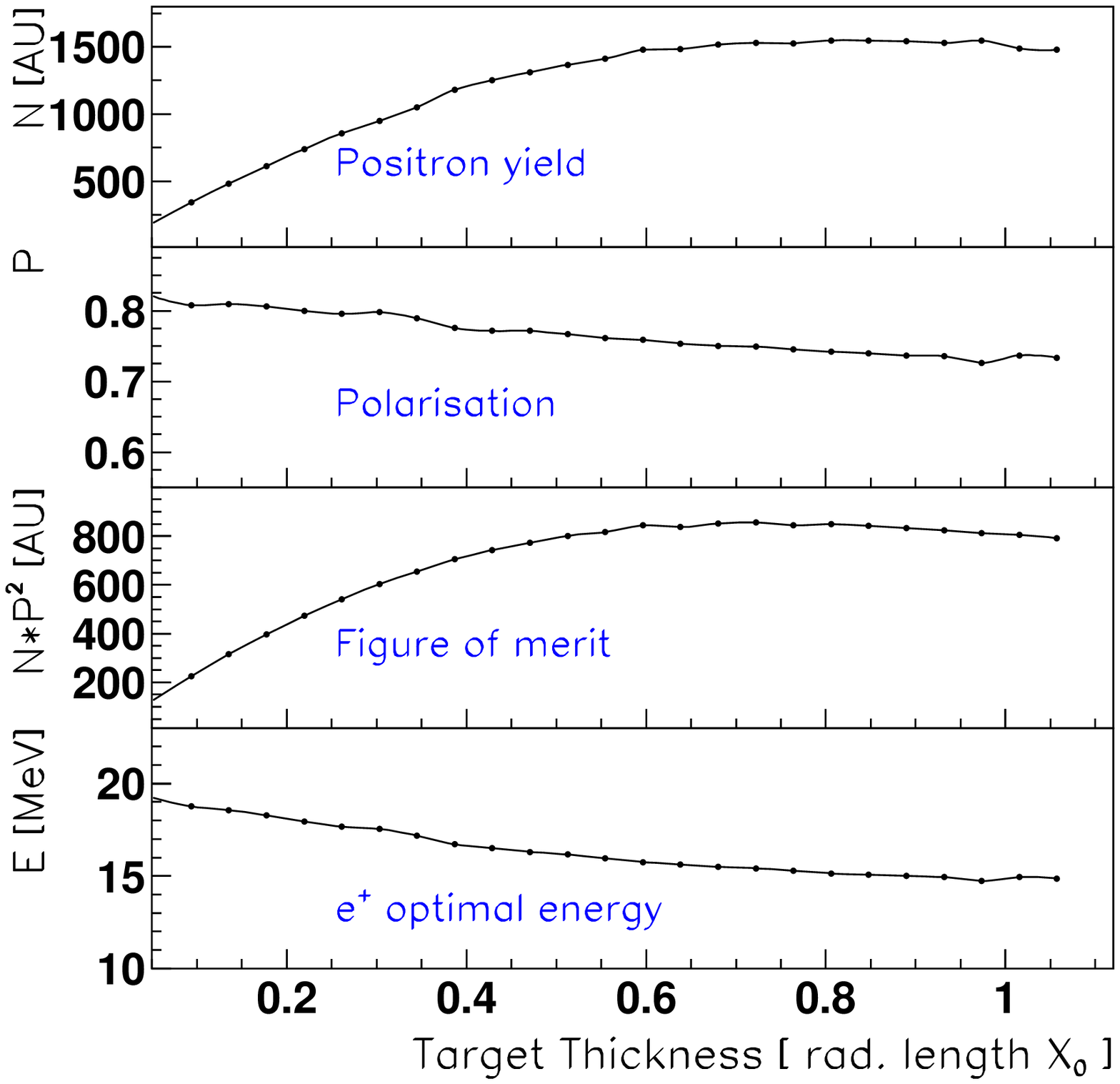}
\caption{The same as in Figure~\ref{Fig:4} for the laser produced positrons.}
\label{Fig:5}
\end{figure}

To choose best production target we try tungsten and titanium changing
their thickness by $0.04X0$ steps, each time recording the positrons
yield, polarisation and energy at the maximum figure of merit. Resulting 
numbers are displayed on Figure~\ref{Fig:4} and Figure~\ref{Fig:5} 
for the undulator and laser case respectively.  

One can note that in general the positron polarisation depends weekly 
on the target thickness i.e. the target optimization could be done by 
maximizing only the positron yield. The distributions also indicate that 
the tungsten is preferable with a thickness of $0.3X_0$ for the undulator 
and $0.7X_0$ for the laser case.

\section{Summary}
In addition to the existing MC programs GEANT-3 is modified to count the 
polarisation. For energies lower than the critical, calculation errors could be large, 
special attention deserve multiple scattering and continuous energy loss.

For the target choice polarized calculations could almost be escaped, its sufficient to maximize
the positron yield. 

\begin{footnotesize}

\end{footnotesize}

\end{document}